\documentclass[aps,prl,reprint]{revtex4-1}
\usepackage{CJK}
\usepackage{graphicx}				
\usepackage{commath,color,soul,enumitem}
\soulregister\cite7
\soulregister\ref7
\usepackage{siunitx} 
\setlist{itemsep=0pt}

\begin{document}
\begin{CJK*}{UTF8}{}
\title{Liquid-grain mixing suppresses droplet spreading and splashing during impact}
\author{Song-Chuan Zhao \CJKfamily{gbsn}(赵松川)}
\email[]{s.zhao@utwente.nl}
\author{Rianne de Jong}
\author{Devaraj van der Meer}

\affiliation{Physics of Fluids Group, Faculty of Science and Technology, University of Twente, PO Box 217, 7500 AE Enschede, The Netherlands}

\date{\today}

\begin{abstract}
Would a raindrop impacting on a coarse beach behave differently from that impacting on a desert of fine sand? We study this question by a series of model experiments, where the packing density of the granular target, the wettability of individual grains, the grain size, the impacting liquid, and the impact speed are varied. We find that by increasing the grain size and/or the wettability of individual grains the maximum droplet spreading undergoes a transition from a capillary regime towards a viscous regime, and splashing is suppressed. The liquid-grain mixing is {discovered to be} the underlying mechanism. An effective viscosity is defined accordingly to quantitatively explain the observations.
\end{abstract}


\maketitle
\end{CJK*}

\textit{Introduction.-- }Droplet impact has been studied over a century since the spark visualizations of Worthington~\cite{worthington1876,*worthington1908}. {Owing to the development of experimental techniques and computation power, our knowledge about the dynamics of droplet impact upon a solid surface or a liquid pool has 
greatly improved~\cite{Rein1993,*Yarin2006,*Josserand2016}.} In general, the dynamics, quantified by, e.g., 
the maximum spreading diameter and the splashing threshold, are governed by the interplay of three forces, namely those due to viscosity, surface tension, and inertia of the impacting droplet. In accordance with which forces are dominant, two distinct regimes can be identified~\cite{CLANET2004,Mundo1995}. 

In contrast, and despite of its ubiquity, droplet impact on sand did not attract much attention until recently~\cite{Katsuragi2010,Marston2010,Delon2011,Nefzaoui2012,Marston2012,Zhao2015a,Zhao2015}, and the underlying physics is still largely unexplored. There are at least two unique features about droplet impact on sand. One is the particular force response of a granular target which can be both solid-like and liquid-like~\cite{Jaeger1996}. The other is the possibility of mixing between liquid and grains which has been shown to be responsible to the formation of various crater morphologies~\cite{Katsuragi2010,Delon2011,Zhao2015,Zhao2015a}. These features add new dimensions to the {parameter space} of droplet impact phenomena, \textit{e.g.}, the properties of individual grains and the whole packing, and therefore present new challenges as well. Besides  potential applications in environmental science and agriculture~\cite{Joung2015}, revealing the role that these new parameters play provides a framework to test to what extent the  concepts established for the conventional droplet impact phenomena may be applied. In this paper, we {report our experimental study of the effect of the wettability of individual grains and the grain size on droplet impact dynamics.}

\textit{Experimental Methods.-- }In our experiments the impacting droplet is composed of either water or ethanol mixed with food dye (mass fraction $< 2\%$) for visualization purposes. The diameter of the water droplets, $D_0$, is fixed to \SI{2.8}{mm} for most experiments and to \SI{3.5}{mm} occasionally. The diameter of the ethanol droplets is in general fixed to \SI{1.8}{mm} and to \SI{2.5}{mm} occasionally. The impacting droplet is released from a nozzle above the substrate. The impact speed, $U$, reaches from  $\SI{1.1}{m/s}$ to $\SI{5.5}{m/s}$ by altering the falling height. The target consists of a bed of beads which is prepared at a packing density in the range of $\numrange[range-phrase = -]{0.55}{0.63}$ by air fluidization and taps. While the droplet deformation is visualized with a high-speed camera, at the same instance the deformation of the substrate surface is measured by an in-house built high-speed laser profilometer~\cite{Zhao2015}.

 We used three types of wettabilities for beads of various sizes [cf.~Table~\ref{t.grains}]: hydrophobic silane coated soda-lime, hydrophilic $\mathrm{ZrO_2}$ ceramic, and {very hydrophilic $\mathrm{ZrO_2}$ ceramic cleaned with a piranha solution}. The grain size, $d_g$, is represented by the mean of the size distribution which is measured under a microscope for a sample of more than 100 grains. The contact angle of both types of ceramic beads is measured by recording the penetration time after a droplet deposition on a packing of grains~\cite{Denesuk1993}, and no penetration is observed for the silane coated beads.
 
 \begin{table}
 \centering
 \caption{\label{t.grains} 
 Contact angles for water and ethanol, $\theta_w$ and $\theta_e$, and 
 grain size $d_g$ for the used granular materials.
 } 
 \begin{ruledtabular}
 \begin{tabular}{lccc}
 material & $d_{g}$ [\si{\micro m}] & $\cos\theta_w$ & $\cos\theta_e$ \\ \hline
 silane coated soda-lime & 114, 200 & $<0$ & - \\
 ceramic  & 98, 167, 257 & 0.3 & - \\
 piranha-cleaned ceramic  & 98, 167, 257 & 0.6-0.7\footnote{Due to aging under exposure to the ambient air, the contact angle of cleaned ceramic beads varies, however, its value is measured after the experiments of each dataset.} & 1 \\
 \end{tabular}
 \end{ruledtabular}%
 \end{table}

\textit{Maximum droplet spreading.-- }It is well known that the rigidity of a granular substrate is very sensitive to its packing density, $\phi$~\cite{Schroter2007,*Gravish2010,*Metayer2011,Umbanhowar2010}. In a previous paper we have discussed the dependence of the maximum droplet spreading diameter, $D_{m}$, on $\phi$~\cite{Zhao2015} and have shown that it can be understood from the partition of the kinetic energy of the impacting droplet into the deformation of both droplet and the substrate. This partition leads to replacing the Weber number, $\textrm{We}=\tfrac{\rho_lD_0U^2}{\sigma}$, which is used to describe droplet spreading when it is limited by surface tension $\sigma$, by an effective Weber number, $\textrm{We}^\dagger=\tfrac{D_0}{D_0+2Z_m}\textrm{We}$, where $Z_m$ is the maximum vertical deformation of the substrate measured by the dynamic laser profilometry and $\rho_l$ is the liquid density. It has been shown that $\textrm{We}^\dagger$ collapses the $D_m$ data for various packing densities~\cite{Zhao2015}.

\begin{figure}
\centering
\includegraphics[width = 8cm]{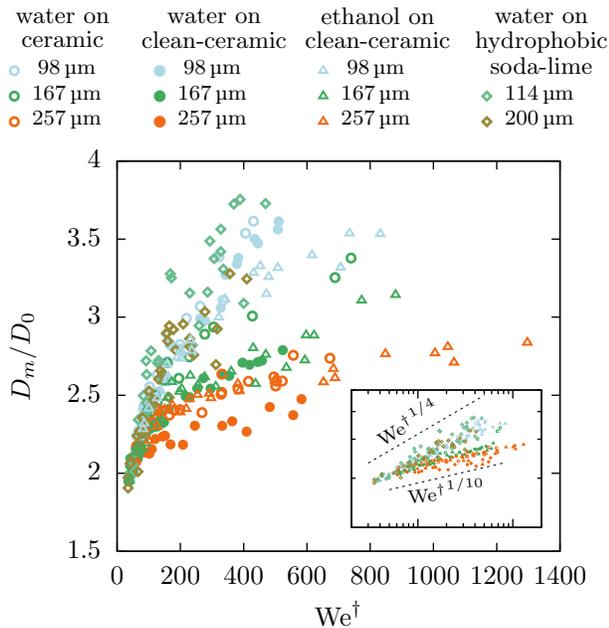}
\caption{\label{f.dm_norm} Maximum droplet spreading diameter $D_{m}$ scaled by the initial diameter of the droplet $D_0$ versus the effective Weber number $\textrm{We}^\dagger$ (see the text for its definition). The results are plotted for different grain sizes (indicated by colors) and combinations of droplets and granular substrates (denoted by symbols). Inset shows the same data in logarithmic scale. 
}
\end{figure}

In Fig.~\ref{f.dm_norm}, $D_m$ normalized by $D_0$ is plotted against the effective Weber number $\textrm{We}^\dagger$ for various combinations of liquids, grain types and grain sizes. It comes as no surprise to see that $D_m$ increases with $\textrm{We}^\dagger$, yet the large spread in Fig.~\ref{f.dm_norm} clearly indicates that $\textrm{We}^\dagger$ alone is not sufficient to describe droplet spreading. Taking a closer look at the dataset, four features can be distinguished:
\begin{description}[before={\renewcommand\makelabel[1]{\bfseries {##1}}}]
\item[{\romannumeral 1})]  The spreading diameter $D_m$ is suppressed with increasing grain size for any given combination of liquid and hydrophilic grain type
(circles and triangles in the figure)
\item[{\romannumeral 2})] For 
hydrophobic soda-lime beads, the grain size does not significantly affect $D_m$ (open diamonds);
\item[{\romannumeral 3})] Water droplets impacting on the very hydrophilic ceramic grains result in smaller $D_m$ than those impacting on plain ceramic grains (open and solid circles); 
\item[{\romannumeral 4})] When plotted in doubly logarithmic scale, the data appears to separate along two power laws: ${\textrm{We}^{\dagger}}^{{1}/{4}}$ and ${\textrm{We}^\dagger}^{1/10}$ (inset).
\end{description}
To summarize, these features indicate that the bulk wettability of the substrate affects $D_m$. This bulk wettability {contains both} the permeability of the substrate and the wettability of individual grains. The crucial question is therefore: how does the bulk wettability influence the relation between $D_m$ and $\textrm{We}^\dagger$? Our investigation begins with a clue provided by the last listed feature.

The two different power laws observed in the inset of Fig.~\ref{f.dm_norm} imply different stopping mechanisms for droplet spreading. The impacts on hydrophobic grains and those on small hydrophilic grains behave as $D_m/D_0\propto{\textrm{We}^{\dagger}}^{1/4}$ which indicates a force balance between inertia and surface tension~\cite{CLANET2004,Zhao2015}. However, for the impacts on large hydrophilic grains we observe another type of scaling, namely close to ${\textrm{We}^{\dagger}}^{1/10}$. Such behavior is equivalent to $D_m/D_0\propto U^{1/5}\propto \textrm{Re}^{1/5}$ which is a hallmark of the dominance of viscous dissipation~\cite{Eggers2010,CLANET2004}, where the Reynolds number, $\textrm{Re}=UD_0/\nu_l$, stands for the significance of inertia relative to viscosity.

For a droplet impacting on a solid surface, the scaling $D_m/D_0\propto\textrm{Re}^{1/5}$ can be understood as follows. While the droplet flattens during spreading, the thickness of the viscous boundary layer grows with time like $\sim\sqrt{\nu_l t}$, where $\nu_l$ is the kinematic viscosity of the liquid. If at the moment of maximum spreading  the thickness of the liquid film, $\sim D_0^3/D_m^2$, matches that of the boundary layer, $\sqrt{\nu_l D_m/U}$, the spreading flow is stopped by viscosity, and one recovers the relation $D_m/D_0\propto\textrm{Re}^{1/5}$~\cite{Eggers2010}. It is plausible that the spread in Fig.~\ref{f.dm_norm} may be interpreted as a transition from a capillary regime to a viscous one. However, since the liquid viscosity, $\nu_l$, is virtually constant for all studied impacts, it is clear that the Reynolds number of the droplet is insufficient to explain such a transition. Nonetheless, the effect of the bulk wettability observed in Fig.~\ref{f.dm_norm} inspired us to regard the mixing between liquid and grains as a boundary layer. In analogy to the viscous boundary layer, this mixing layer ceases liquid motion within it, due to strong viscous dissipation at the length scale of a grain. For hydrophobic grains the mixing is negligible, which explains that for those grains no grain size dependence of $D_m$ is observed. However, for hydrophilic grains the droplet spreading dynamics 
may well be altered. Therefore, to understand the two power laws shown in Fig.~\ref{f.dm_norm} we analyze the development of the mixing layer.

\textit{Effective viscosity.-- }We use Darcy's law to quantify the penetration flux of the impacting {droplet} into the substrate,
\begin{equation}
\vec{Q}= \frac{\kappa A}{\mu_l}\nabla{P}.
\label{eq.darcy}
\end{equation}
In the above equation the permeability of the substrate, $\kappa=(1-\phi)^3d_g^2/(180\phi^2)$, is  defined by the Carman-Kozeny relation~\cite{Carman1956}, $\nabla P$ is the pressure gradient, $A$ is the contact area between the droplet and the substrate, and $\mu_l=\rho_l\nu_l$ is the dynamic viscosity of the liquid. {Since the pressure gradient is mainly in the vertical direction, Eq.~({\ref{eq.darcy})} can be reduced to a scalar equation. The penetration of liquid into the substrate can now be viewed as the growth of a `boundary'  layer into the droplet, whose thickness, $L$, is defined by its time derivative: ${\dif L}/{\dif t}=Q/A$. $L$ denotes the thickness of the liquid layer that merges with the sand, but due to the presence of the grains the penetration depth of the liquid into the sand bed is larger, namely $L/(1-\phi)$, 
and the pressure gradient can be estimated as $(1-\phi)P/L$. Eq.~(\ref{eq.darcy}) thus becomes an ordinary differential equation for the mixing layer thickness $L$ with respect to time $t$, and its solution is
\begin{equation}
L(t) = \sqrt{\frac{2\kappa P (1-\phi)}{\mu_l}\,t}.
\label{eq.pen_depth}
\end{equation}

Besides the aforementioned physical analogy between the mixing layer and the viscous boundary layer, Eq.~(\ref{eq.pen_depth}) indicates that the analogy extends to the mathematical form of the growth of their thicknesses as well, \textit{i.e.,} both are diffusive. Therefore, it can be used to define an effective viscosity, the quantity $\nu_p\equiv 2\kappa P(1-\phi)/\mu_l$ that appears in front of $t$. While most quantities in Eq.~(\ref{eq.pen_depth}) are merely properties of the substrate or the impacting liquid, the pressure, $P$, that drives mixing is not. Therefore, estimating $P$ is the last remaining piece of the puzzle.

There are three potential sources of the driving pressure $P$: inertia, capillarity and gravity. We estimate their order of magnitudes with typical parameters for the water droplets used in our experiments: liquid density $\rho_l=\SI{1.0d3}{kg/m^3}$,  surface tension $\sigma=\SI{72d-3}{N/m}$, impact speed $U\sim \numrange[range-phrase = -]{1}{5}\,\si{m/s}$, droplet diameter $D_0\approx \SI{3}{mm}$, and grain size $d_g\sim \SI{100}{\micro m}$. Then one obtains a typical inertial pressure of $P_{i}\approx\rho_l U^2\sim 10^3-\SI{d4}{Pa}$, a capillary pressure of $P_c \approx 4\sigma\cos\theta_c/d_g\sim 10^{3}\cos\theta_c\,\si{Pa}$, and a gravitational pressure of $P_g \approx \rho_l g D_0\sim \SI{10}{Pa}$. For the liquids and hydrophilic grains that we used the contact angle stays in a range of $\cos\theta_c\in[0.3,\,1]$, hence, $P_c$ is at least one order of magnitude larger than $P_g$ which is therefore neglected. Though $P_i$ is again at least one order of magnitude larger than $P_c$, previous simulation and experimental works have shown that $P_i$ only acts within an inertial time scale $\tau_i\approx D_0/U$~\cite{Soto2014,Eggers2010}. We correct this time scale as $\tau_i=(D_0+2Z_m)/U$ by taking the deformation of the substrate, $Z_m$, into account. In contrast, $P_c$ lasts as long as the contact between liquid and grains exists. {This contact time is estimated as half of the intrinsic oscillation time of the droplet~\cite{Delon2011,Richard2002,*Okumura2003}, 
$\tau_c=\tfrac{1}{2}\sqrt{\tfrac{\pi}{6}\tfrac{\rho_l {D_0^3}}{\sigma}}$, 
and represents the time it takes
until maximum droplet spreading is reached. Note that in general $\tau_c>\tau_i$. These two time scales provide relative weights for $P_i$ and $P_c$ in the spreading phase of the droplet, and the average effect of the total pressure is evaluated as $P = \tfrac{\tau_i}{\tau_c}P_i+P_c$~\cite{Suppl}.} Inserting this total pressure into Eq.~(\ref{eq.pen_depth}), the effective viscosity is estimated as
\begin{equation}
\nu_p = \frac{2\kappa (1-\phi)}{\mu_l}P=\frac{2\kappa (1-\phi)}{\mu_l}\left(\frac{\tau_i}{\tau_c}P_i+P_c\right),
\label{eq.eff_v}
\end{equation}
and a corresponding effective Reynolds number, ${\textrm{Re}^\dagger=\tfrac{UD_0}{\nu_p}}$, is defined.

When evaluating $\nu_p$, the inertial pressure (as in our previous study~\cite{Zhao2015}) is corrected by the deformation of the substrate $Z_m$, $P_i = \rho_l U^2\tfrac{D_0}{D_0+2Z_m}$; the capillary pressure is given by $P_c=4\sigma\cos\theta_c/d_c$, where $d_c=\tfrac{2(1-\phi)}{3\phi}d_g$ is the average diameter of capillaries between grains derived from the Carman-Kozeny relation; and a characteristic packing density $\phi^*=0.59$ is used for all packings during impact~\cite{Umbanhowar2010,Zhao2015}. We then find that $\nu_p$ is in the range of \numrange[range-phrase = --]{e-5}{e-4} \si{m^2/s}, \textit{i.e.}, at least one order of magnitude larger than the kinematic viscosity of water~\footnote{Note that as a consequence the viscous boundary layer inside the droplet can be neglected.}. It is worthy to point out that $\nu_p$ is inversely proportional to liquid viscosity $\nu_l$, and therefore when using highly viscous liquids the viscous boundary layer is likely to become dominant.

\begin{figure}
\centering
\includegraphics[width = 8cm]{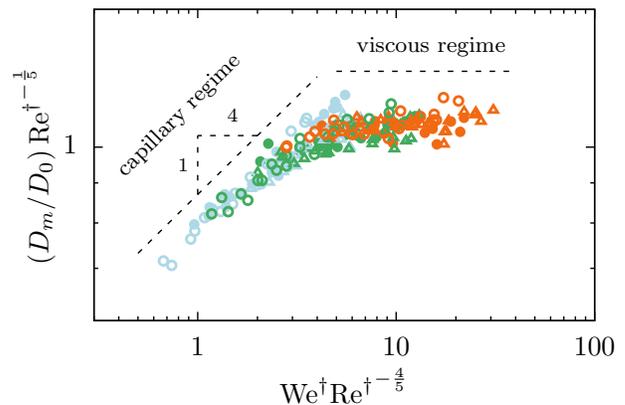}
\caption{\label{f.dm_Re} The maximum droplet spreading diameter, $D_m/D_0$, for all hydrophilic impacts of Fig.~\ref{f.dm_norm} in a doubly logarithmic plot. The same symbols and colors as in Fig.~\ref{f.dm_norm} are used. The data have been compensated in such a way that a transition between a capillary ($\propto{\textrm{We}^\dagger}^{{1}/{4}}$) and a viscous regime ($\propto{\textrm{Re}^\dagger}^{1/5}$) can be observed. The power laws of these two regimes are indicated by dashed lines.}
\end{figure}

The effective viscosity defined in Eq.~(\ref{eq.eff_v}) grows with increasing grain size, $d_g$, on which it depends through $\kappa$ and $P_c$. In consequence, for large $d_g$ the droplet spreading is more likely to be stopped by liquid-grain mixing before surface tension can do so, and hence $D_m/D_0\propto{\textrm{Re}^\dagger}^{{1}/{5}}$ would be expected. In contrast, for small $d_g$ mixing is slower and the surface tension balances inertia, leading to $D_m/D_0\propto{\textrm{We}^\dagger}^{1/4}$. To illustrate the transition between these two scaling relations, data of all hydrophilic impacts are plotted as ${D_m}/{D_0}\,{\textrm{Re}^\dagger}^{-{1}/{5}}$ versus $\textrm{We}^\dagger{\textrm{Re}^\dagger}^{-4/5}$ in Fig.~\ref{f.dm_Re}~\cite{CLANET2004}. The newly introduced $\textrm{Re}^\dagger$ successfully collapses data of various surface tensions, grain sizes, and wettabilities on a master curve without free parameters. Further discussion on the scaling laws can be found in Supplementary material~\cite{Suppl}.

Leaving the mathematical details aside here~\cite{Suppl}, the transition in Fig.~\ref{f.dm_Re} can be interpreted as a crossover from a regime where $D_0$ is the dominant length scale to one where both $D_0$ and $d_g$ matter, which, since $d_g\ll D_0$, implies that viscous dissipation in the mixing layer becomes important. This happens when $\nu_p$ is large, \textit{i.e.,} $\textrm{Re}^\dagger$ is small. Previous studies about droplet spreading on sand have used the traditional Weber number and reported various scaling relations~\cite{Marston2010,Marston2012,Nefzaoui2012}. The introduction of $\mathrm{We}^{\dagger}$ and $\mathrm{Re}^{\dagger}$ [cf. Fig.~\ref{f.dm_Re}], which take the deformability and bulk wettability of the substrate into account respectively, may {provide a universal framework to understand droplet spreading when impacting on sand or other porous media.}

\textit{Splashing suppression.-- }With increasing impact velocity the inertia of the spreading liquid may overcome both surface tension and viscosity, and splashing can occur. Therefore, for impact of droplets on solid substrates at a given Weber number, the Reynolds number determines whether a droplet will splash or not~\cite{Mundo1995}. Is the same true for the effective Reynolds number, $\textrm{Re}^\dagger$, introduced here? {As the effective viscosity $\nu_p$ increases with ${d_g}$, resulting in a smaller $\textrm{Re}^\dagger$, large grains are expected to suppress the splash.} Indeed, as shown in Fig.~\ref{f.splash},   an ethanol droplet already splashes for $\textrm{We}^\dagger=431$ when impacting on ceramic beads of $d_g=\SI{98}{\micro m}$, whereas when impacting on the same grain type but with $d_g=\SI{257}{\micro m}$, splashing is delayed until $\textrm{We}^\dagger>652$~\footnote{The splashing considered here consists of fragments expelled from a fully developed droplet rim rather than the prompt splashing which happens at the very early stage of the impact.}. To quantify the splashing threshold, Mundo \textit{et al.}~\cite{Mundo1995} proposed a dimensionless splashing parameter, $K_d=\textrm{We}^{{1}/{2}}\textrm{Re}^{{1}/{4}}$ relating inertial force to viscous and surface tension forces. Here, we replace the Weber and Reynolds number by their effective counterparts in the definition of $K_d$, which leads to $K_d^\dagger = {\textrm{We}^\dagger}^{1/2}{\textrm{Re}^\dagger}^{1/4}$. A transition can be seen around $K_d^{\dagger}\approx 85$ for all hydrophilic impacts in Fig.~\ref{f.splash}. It is necessary to point out that since the definition of $K_d$ is insensitive to substrate properties such as wettability and roughness~\cite{Range1998,*Rioboo2001}, the value of the splashing threshold differs from one situation to another, \textit{e.g.}, different values of $K_d = 57.7,\, 80,$ and 120 are reported for impacts on a solid surface~\cite{Mundo1995}, nanofibers~\cite{Lembach2010}, and dry granular packings~\cite{Nefzaoui2012} respectively. Therefore, the threshold value reported here is not intended to be compared directly with the above mentioned ones. Nevertheless, the existence of a unified splashing threshold for impacts on different grain sizes is another manifestation of 
how liquid-grain mixing 
is captured by $\textrm{Re}^\dagger$.

\begin{figure}
\centering
\includegraphics[width = 8cm]{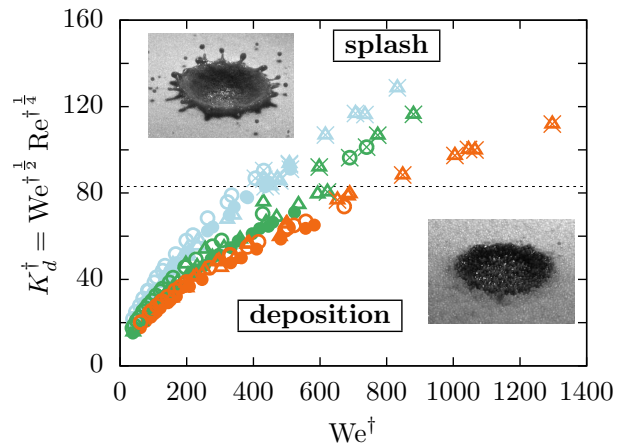}
\caption{\label{f.splash} The splashing parameter $K_d={\textrm{We}^\dagger}^{1/2}{\textrm{Re}^\dagger}^{1/4}$ as a function of $\textrm{We}^\dagger$. Here, $\textrm{We}^\dagger$ and $\textrm{Re}^\dagger$ are the effective Weber and Reynolds number, as defined in the text. The same colors and symbols as in Fig.~\ref{f.dm_norm} are used to denote various combinations of liquid and wettability of grains, while those impacts resulting in splashing/fragmentation are highlighted with a cross. The dashed line marks the threshold separating deposition and splashing regimes. The top left and bottom right insets show examples of a water droplet with $D_0=\SI{2.8}{mm}$ and $U=\SI{5.2}{m/s}$ impacting on clean ceramic beads with $d_g=\SI{98}{\micro m}$ and ceramic beads with $d_g=257\si{\micro m}$ respectively.}
\end{figure}

\textit{Discussion.-- }In this paper we introduced effective Weber and Reynolds numbers $\textrm{We}^{\dagger}$ and $\textrm{Re}^{\dagger}$, which incorporate the deformability and bulk wettability of a granular substrate respectively. This reveals the hidden similarities between droplet impact on sand and that on a solid substrate for two aspects:  
maximum droplet spreading and splashing. 
Despite of the similarities represented by $\textrm{We}^{\dagger}$ and $\textrm{Re}^{\dagger}$, there are distinctions resulting from the characteristics of a sand bed. One example stems from the mobility of individual dry grains which can result in a shear band under external driving~\cite{Jaeger1996}. It is thus plausible that, when mixing between liquid and grains is subtle, the boundary condition experienced by a spreading droplet on sand is neither purely slip nor no-slip but one with a finite slip length~\cite{Brochard1992} with the magnitude of the grain size. Another example is the role of ambient air. Owing to recent development of high-speed imaging techniques, ambient air is found to be responsible for splashing~\cite{Xu2005,*Riboux2014} and bubble entrapment~\cite{Driscoll2011,*Mani2010,*Bouwhuis2012,*Ruiter2012}. In contrast, the permeability of a sand bed may prevent the existence of such a thin air film. This also differentiates splashing suppression in Fig.~\ref{f.splash} from that on deformable substrates~\cite{Howland2015}. Further work is necessary to understand the role of these unique features of a sand bed on the impact dynamics.

This work is financed by the Netherlands Organisation for Scientific Research (NWO) through a VIDI Grant No. 68047512.

\bibliography{drop_impact}

\end{document}